\documentstyle{article}
\newcommand{\al}{\alpha}
\newcommand{\Om}{\Omega}
\newcommand{\af}{ATFT}
\newcommand{\cic}{CIBC}

\newcommand{\gf}{generating function}
\newcommand{\k}{K_{\pm}}
\newcommand{\st}{\stackrel}
\newcommand{\de}{\delta}
\newcommand{\kog}{\quad\quad}
\newcommand{\wt}{\widetilde}

\newcommand{\be}{\beta}
\newcommand{\ii}{\int \limits_{-\infty}^{+\infty}}

\newcommand{\th}{\theta}
\newcommand{\ot}[2]{\left\{#1 \stackrel{\otimes}{,} #2\right\}}
\newcommand{\la}{\lambda}

\newcommand{\ptl}{\partial}
\newcommand{\ga}{\gamma}

\def\nn{ \nonumber }
\def\ll{ \label }
\def\frac#1#2{{#1\over #2}}

\newenvironment{eq}{\begin{equation}}{\end{equation}}
\newenvironment{ea}{\begin{eqnarray}}{\end{eqnarray}}
\newenvironment{ea*}{\begin{eqnarray*}}{\end{eqnarray*}}
\begin{document}

\begin{center}
 {\Large \bf Classically and Quantum Integrable Systems} \\
\vspace{3mm} 
{\Large \bf with Boundary}  \\
\vspace{5mm}
 {\large Yi-Xin Chen$^{\dag}$,  Xu-Dong Luo$^{\ddag}$  and Ke Wu$^{\ddag}$} \\   
 
\vspace{3mm}
\begin{minipage}{15cm}
  \dag  \quad Institute of Zhejiang Modern Physics and Department of Physics, 

     \indent\quad $\;$ Zhejiang University, $\;\;$ Hangzhou 310027, $\;\;$ China 

  \ddag \quad Institute of Theoretical Physics, \quad Academia Sinica, $\;\;$ Beijing 100080, China 

\vspace{6mm}
\centerline{\bf Abstract}  

\vspace{5mm}
\quad
 We study two-dimensional classically integrable field theory with independent
 boundary condition on each end, and obtain three possible generating functions
 for integrals of motion when this model is an ultralocal one. Classically
 integrable boundary condition can be found in solving boundary $\k$ equations.
 In quantum case, we also find that unitarity condition of quantum $R$- matrix
 is sufficient to construct commutative quantities with boundary, and its
 reflection equations are obtained.
\end{minipage}
\end{center}

\section{Introduction}
\indent
  Recently, there has been great progress in understanding two-dimensional
integrable field theory on a finite interval with independent boundary condition
on each end \cite{C}-\cite{BCD}. The motivation is not only from the necessary
of itself, but also from the studies in boundary-related phenomena in statistical
systems near criticality \cite{B} and integrable deformations of conformal field
theories \cite{CL}.   \\
 \indent  
In order to deal with integrable models with boundary, relying on previous
results of Cherednik \cite{C}, Sklyanin \cite{{S},{S1}} introduced a new generating
function which originates from the periodic boundary one. In classically integrable
models \cite{S}, if there has a well-known relation for monodramy matrix \cite{FT}
as $\ot{T}{T}=[r,\; T\otimes T]$, and $r$-matrix satisfies the condition of
$r(\al)=-r(-\al)$, the new generating function defined in $[x_{-},x_{+}]$ can
be expressed as:
\begin{eq}
   \tau(\al) \equiv tr\left\{ K_{+}(\al) T(x_{+},x_{-},t,\al)
         K_{-}(\al) T^{-1}(x_{+},x_{-},t,-\al) \right\}  , \ll{sk-tau} 
\end{eq}
where $\k$ are boundary reflection matrices.  \\
\indent
Expanded as a Laurent series in $\al$, all coefficients of $\tau(\al)$ make an
infinite number of integrals of motion which ensure the completely integrability
of the model. From \cite{FT}, $\tau(\al)$ must be in involution between different
spectral parameter $\al$ and it is independent of time. In other words, $\k$
must satisfy some constraint equations, and existence of nontrivial $\k$ solutions
means there are nontrival classically integrable boundary conditions (CIBC).   \\
\indent
There has no condition of $r(\al)=-r(-\al)$ in affine Toda field theory (\af),
so P.Bowcock {\it et al} \cite{BCD} developed a method of modified Lax pair to
deal with such models, in which the new generating function in $(-\infty,x_{+}]$
reads
\begin{eq}
   \tau(\al) \equiv tr\left\{  T^{\dag}(-\infty,x_{+},t,-\al) K_{+}(\al)
    T(-\infty,x_{+},t,\al)  \right\}    , \ll{b-tau}
\end{eq}
in which $"\dag"$ denotes conjugation and it has little difference with the
original paper\cite{BCD} according to different definition in $T$ matrix. We
must point out that boundary Lax pair in \cite{BCD} has been modified from the
periodic boundary one. \\
\indent
In this paper, we find it is necessary to add a new parameter to the generating
function (\ref{sk-tau}) in order to deal with {\af}, and no symmetry condition
of $r$- matrix is needed in fact. Besides this modified form, we also construct
two other possible generating functions by zero curvature representation.
After we extend our results to quantum integrable systems, we find unitarity
condition of quantum $R$- matrix is sufficient to construct commutative
quantities
with boundary too.  \\
\indent
 The paper is organized as follows. In section 2, three possible generating
functions are constructed by zero curvature representation. In order to regard
constructed functions as generating functions, algebra equations (reflection
equations) and evolution equations of $K_{\pm}$ matrices appear in section 3.
Then, we study {\af} in section 4 and find the links among these generating
functions. In section 5, we extend our results to the quantum case, and demonstrate
that unitarity condition of quantum $R$- matrix is sufficient to construct
commutative quantities. Then, we compare our commutative quantities with those
of paper \cite{MN} and find the relation between them. At last, some discussion
will be found in section 6.

\section{Construction of generating function}
\vspace{-5mm}
\subsection{Periodic boundary condition}
\vspace{-5mm}
 The zero curvature approach to inverse scattering \cite{FT} relies on the existence of a pair of linear partial differential 
equations in $d\times d$ matrix. 
  $$
       {\ptl_{x}}\Psi  = U(x,t,\al)\Psi     , \quad \;
       {\ptl_{t}}\Psi  = V(x,t,\al)\Psi    ,
  $$
here, Lax pair $(U,V)$ are $d\times d$ matrices whose elements are functions of complex valued field $\phi(x,t)$ and its derivatives, 
$\al\in C$ is a spectral parameter. Zero curvature condition appears from
compatibility of the above equation, it is
\begin{eq}
      {\ptl_{t} U}-{\ptl_{x} V}+{\left[ U, \; V\right]}=0    .
\end{eq}
By zero curvature representation, we define transition matrix
\begin{eq}
   T(x,y,t,\al)={\cal P} \, exp\left\{\int_{y}^{x}U(x^{\prime},t,\al)dx^{\prime} \right\}  , \;\;  x\geq y ,
\end{eq}
where ${\cal P}$ denotes a path ordering of non-commuting factors. Now, $T$ matrix satisfies
\begin{eqnarray}
       {\ptl_{x}}T & =& U(x,t,\al) T   , \nn \\
       {\ptl_{t}}T & =& V(x,t,\al) T -T V(y,t,\al)  ,\ll{tm} \\
        Id &=&  T(x,x,t,\al)   , \nn
\end{eqnarray}
where $Id$ is $d\times d$ identity matrix. \\
\indent
It is well known that trace of monodramy matrix $T_{L}(t,\al)\equiv T(L,-L,t,\al)$ is a generating function on periodic boundary 
condition, so is another more explicit form $\tau(\al)=\ln tr T_{L}(t,\al)$. Expanded as a Laurent series in $\al$, $\tau(\al)$ make 
an infinite number of integrals of motion. Conservation condition of these
integrals can be proved by the second equation of (\ref{tm})
with periodic boundary condition, and the involution condition is proved in Poisson bracket:
\begin{eq}
  \ot{T(x,y,\al)}{T(x,y,\be)}=[r(\al,\be),T(x,y,\al)\otimes T(x,y,\be)] \quad ,L\geq x\geq y\geq -L    ,
\end{eq}
in which, $T$ is a $d\times d$ matrix, $\st{1}{T}\equiv T\otimes Id$ and $\st{2}{T}\equiv Id\otimes T$.
$r(\al,\be)$ is a $d^{2}\times d^{2}$ matrix whose elements depend on $\al$ and $\be$ only. The Jacobi identity 
for the bracket holds if and only if $r$-matrix is a solution of classical Yang-Baxter equation. \\ 
\indent
 In periodic boundary condition, it is obvious that $\ot{\tau(\al)}{\tau(\be)}=0$. 
So $\tau(\al)$ constructs a family of generating function for integrals of motion.

\vspace{-5mm}
\subsection{Independent boundary condition}
\vspace{-5mm}
  As soon as periodic boundary condition is broken, $\tau(\al)$ defined before should be not a conservative quantity, so that we have to find a new expression of
generating function.
As discussed in papers \cite{{G&Z},{BCD}}, if Lagrangian density in bulk theory is ${\cal L}_{f}$, then,
the new Lagrangian density with boundary appears as: 
\begin{eq}
  {\cal L}=\th(x_{+}-x)\th(x-x_{-}) {\cal L}_{f}
           -\de(x_{+}-x) V_{+}(\phi(x_{+}),\ptl_{\mu}\phi(x_{+}))
           -\de(x-x_{-}) V_{-}(\phi(x_{-}),\ptl_{\mu}\phi(x_{-}))  . \ll{Lb} 
\end{eq}
By means of principle of the least action associated with (\ref{Lb}), we will obtain motion equation
in $(x_{-},x_{+})$ and boundary equations on each end. \\ 
\indent
If ${\cal L}_{f}$ is expressed as ${\cal L}_{f}=\frac{1}{2} \ptl_{\mu}\phi\ptl^{\mu}\phi-V(\phi)$ and $V_{\pm}$ depend only
on $\phi(x_{\pm})$ (and independent of its derivatives), boundary equations on each end will be
\begin{eq}
  \ptl_{x}\phi=\mp\ptl_{\phi} V_{\pm}  \quad , \quad x=x_{\pm}  .\ll{bc1}
\end{eq}
\indent
On each end, if $\phi(x)$ is a smooth function of coordinate $x$, we can extend the motion equation to whole
domain $[x_{-},x_{+}]$, so are the boundary Lax pair.  In other words, we keep the uniform expression of 
fundamental Poisson bracket in the whole domain, ever if it is on independent boundary condition. \\
\indent 
 Similar to the definition of transition matrix, there is another matrix function
\begin{eq}
 F(x,t,\al)={\cal P}\;\; exp\left\{\int_{t_{0}}^{t}V(x,t^{\prime},\al)dt^{\prime} \right\}  , \; t\geq t_{0}  .
\end{eq}
By zero curvature condition, we construct a quantity which is independent of time:
\begin{eq}
   F^{-1}(x_{+},t_{1},\al)T(x_{+},x_{-},t_{1},\al)F(x_{-},t_{1},\al) 
   =  F^{-1}(x_{+},t_{2},\al)T(x_{+},x_{-},t_{2},\al)F(x_{-},t_{2},\al) .  \ll{qit}
\end{eq}
It can be proved easily because both sides in the equation are equal to $T(x_{+},x_{-},t_{0},\al)$.  \\ 
\indent
With another equation of argument $(-\al+\de)$, which can be obtained by the same method,
we expect generating function with boundary shall be constructed as follows. For example:
\begin{eqnarray*}
 {} & tr  & \left\{ \quad [F(x_{+},t,-\al+\de)F^{-1}(x_{+},t,\al)] \;\; T(x_{+},x_{-},t,\al) \right.  \\
  {} & {} & \left.  \quad \times [F(x_{-},t,\al)F^{-1}(x_{-},t,-\al+\de)] \;\; T^{-1}(x_{+},x_{-},t,-\al+\de)  \right\} . 
\end{eqnarray*}    
\indent 
By its product method, it is obvious that this quantity is a conservative
quantity. If we regard it as a generating function for integrals of motion,
there are the main problem: 1. Does such a constructed quantity satisfy involution condition ?  2. It had better be independent of  
"$t_{0}$" which comes from $F$ matrix. Base on these problems, we introduce $\k$ matrices instead of $F$ terms and impose involution 
and conservation conditions on the new form. The new generating function is \\
\indent $\quad a.$
\begin{eq}
    \tau(\al)=tr\left\{ K_{+}(x_{+},t,\al) T(x_{+},x_{-},t,\al)
         K_{-}(x_{-},t,\al) T^{-1}(x_{+},x_{-},t,-\al+\de) \right\} . \ll{tau1}
\end{eq}
\indent 
Moreover, we use "$\dag$" (conjugation) or "$ t $" (transposition)
instead of "$-1$" (inverse) in order that $K_{+}$ and $K_{-}$ depend only
on the variables of the boundary $x_{+}$ and $x_{-}$, respectively.
The results are: \\
\indent  $\quad b.$
\begin{eq}
    \tau(\al)=tr\left\{ K_{+}(x_{+},t,\al) T(x_{+},x_{-},t,\al)
         K_{-}(x_{-},t,\al) T^{t}(x_{+},x_{-},t,-\al+\de) \right\}  , \ll{tau2}
\end{eq}
\indent  $\quad c.$
\begin{eq}
    \tau(\al)=tr\left\{ K_{+}(x_{+},t,\al) T(x_{+},x_{-},t,\al)
         K_{-}(x_{-},t,\al) T^{\dag}(x_{+},x_{-},t,-\al+\de) \right\}  . \ll{tau3}
\end{eq}
In quantities (\ref{tau1}), (\ref{tau2}) and (\ref{tau3}), $\k$ matrices and $\de$ are similar in symbols only. 
Each of these quantities will be a generating function of integrable systems
with boundary, if both involution and conservation conditions are satisfied.
\section{$\tau(\al)$ as a generating function}
It is well known that generating function for integrals of motion must be in involution (between each other) and independent of time \cite{FT}.
So if we regard quantity (\ref{tau1}) as a generating function, some constraint conditions must be imposed on it. 
Now we study quantity (\ref{tau1}) in this section.  

\vspace{-5mm}
\subsection{Involution condition}
\vspace{-5mm}
Taking the notations similar to paper \cite{S1}, we define:
\begin{eqnarray*}
   {\cal T}_{+}(x,\al)&=&T^{-1}(x_{+},x,t,-\al+\de)K_{+}(x_{+},t,\al)T(x_{+},x,t,\al)  , \\
   {\cal T}_{-}(x,\al)&=&T(x,x_{-},t,\al)K_{-}(x_{-},t,\al)T^{-1}(x,x_{-},t,-\al+\de)  , \\
   {\cal T}(x,\al)&=&{\cal T}_{-}(x,\al) {\cal T}_{+}(x,\al) .
\end{eqnarray*}
Comparing it with quantity (\ref{tau1}), we find $tr {\cal T}(x,\al)$ is just equal to $\tau(\al)$. Now, we impose some constraint
conditions on $\k$ matrices:
 $$ \ot{K_{\pm}(x_{\pm},t,\al)}{T(x_{+},x_{-},t,\be)}=0  , $$
\begin{eq}
  \ot{K_{\pm}(\al)}{K_{\pm}(\be)}=0 \quad , \quad \ot{K_{\pm}(\al)}{K_{\mp}(\be)}=0   . \ll{ck}
\end{eq}
It means
 $$ \ot{{\cal T}_{+}(x,t)}{{\cal T}_{-}(x,t)}=0  . $$
\indent
If $\k$ matrices are independent of field variances,
condition (\ref{ck}) is satisfied naturally. But we must point out
that the $\k$ matrices, in general, can depend on the field variables.
Poisson bracket on ${\cal T}_{\pm}$ is:

\indent
{\bf Proposition 1.} If $K_{+}$ matrix satisfies
\begin{eqnarray}
 0 & =&  -r(-\al+\de,-\ga+\de) {\st{1}{K}}_{+}(t,\al) {\st{2}{K}}_{+}(t,\ga)
    +{\st{1}{K}}_{+}(t,\al) r(\al,-\ga+\de)  {\st{2}{K}}_{+}(t,\ga) \nn \\
 {} & {}  &+ {\st{2}{K}}_{+}(t,\ga) r(-\al+\de,\ga) {\st{1}{K}}_{+}(t,\al)
    -{\st{1}{K}}_{+}(t,\al) {\st{2}{K}}_{+}(t,\ga) r(\al,\ga)   , \ll{ckp1} 
\end{eqnarray}
then, ${\cal T}_{+}$ algebra should obey the following relation:
\begin{eqnarray}
 & & \left\{ \st{1}{\cal T}_{+}(x,\al),\st{2}{\cal T}_{+}(x,\ga) \right\}    \nn \\
 & =&  -r(-\al+\de,-\ga+\de) {\st{1}{\cal T}}_{+}(x,\al) {\st{2}{\cal T}}_{+}(x,\ga)
       + {\st{1}{\cal T}}_{+}(x,\al) r(\al,-\ga+\de)  {\st{2}{\cal T}}_{+}(x,\ga) \nn \\
 & & + {\st{2}{\cal T}}_{+}(x,\ga) r(-\al+\de,\ga) {\st{1}{\cal T}}_{+}(x,\al)
       -{\st{1}{\cal T}}_{+}(x,\al) {\st{2}{\cal T}}_{+}(x,\ga) r(\al,\ga)  . \ll{ckp2}
\end{eqnarray}  

\indent  It should be emphasized that $K_{+}$ is a subalgebra of ${\cal T}_{+}$ algebra according to the definition of ${\cal T}_{+}$.
Proposition 1 can be proved by calculating Poisson bracket on 
${\cal T}_{+}$ directly.
There is another algebra of ${\cal T}_{-}$ similar to ${\cal T}_{+}$:

\indent
{\bf Proposition 2.} If $K_{-}$ matrix satisfies:
\begin{eqnarray}
 0  & =&      r(\al,\ga)    \st{1}{K}_{-}(t,\al) \st{2}{K}_{-}(t,\ga)
   -  \st{1}{K}_{-}(t,\al) r(-\al+\de,\ga)    \st{2}{K}_{-}(t,\ga)       \ll{ckm1} \\
 &{} & -  \st{2}{K}_{-}(t,\ga) r(\al,-\ga+\de)     \st{1}{K}_{-}(t,\al)
   +  \st{1}{K}_{-}(t,\al) \st{2}{K}_{-}(t,\ga) r(-\al+\de,-\ga+\de)    \; , \nn
\end{eqnarray}
then, it leads to the relation of ${\cal T}_{-}$ algebra being:
\begin{eqnarray}
 &{} &  \left\{ \st{1}{\cal T}_{-}(x,\al),\st{2}{\cal T}_{-}(x,\ga) \right\}  \nn \\
 & =&  r(\al,\ga)    \st{1}{\cal T}_{-}(x,\al) \st{2}{\cal T}_{-}(x,\ga)
   -  \st{1}{\cal T}_{-}(x,\al) r(-\al+\de,\ga)    \st{2}{\cal T}_{-}(x,\ga)  \ll{ckm2} \\
 &{} & -  \st{2}{\cal T}_{-}(x,\ga) r(\al,-\ga+\de)     \st{1}{\cal T}_{-}(x,\al)
   +  \st{1}{\cal T}_{-}(x,\al) \st{2}{\cal T}_{-}(x,\ga) r(-\al+\de,-\ga+\de)  .  \nn
\end{eqnarray}  

\indent
 The proof is similar to proposition 1. Used proposition 1 and 2,
   Poisson bracket on ${\cal T}(x,\al)$ can be calculated as follows:
\begin{eqnarray*}
 & &  \left\{ \st{1}{\cal T}(x,\al),\st{2}{\cal T}(x,\ga) \right\} \\
 &=& \left\{ \st{1}{\cal T}_{-}(x,\al) \st{1}{\cal T}_{+}(x,\al)
        , \st{2}{\cal T}_{-}(x,\ga) \st{2}{\cal T}_{+}(x,\ga) \right\} \\
 &=& \st{1}{\cal T}_{-}(x,\al)     \st{2}{\cal T}_{-}(x,\ga)
    \left\{  \st{1}{\cal T}_{+}(x,\al)  ,  \st{2}{\cal T}_{+}(x,\ga) \right\}
 + \left\{  \st{1}{\cal T}_{-}(x,\al) ,    \st{2}{\cal T}_{-}(x,\ga) \right\}
      \st{1}{\cal T}_{+}(x,\al)    \st{2}{\cal T}_{+}(x,\ga)       \\
 &=& \left[r(\al,\ga),\st{1}{\cal T}(x,\al)\st{2}{\cal T}(x,\ga)\right]
    + \left[\st{1}{\cal T}(x,\al) ,\st{2}{\cal T}_{-}(x,\ga) r(\al,-\ga+\de) \st{2}{\cal T}_{+}(x,\ga)\right] \\
 &  & + \left[\st{2}{\cal T}(x,\ga) ,\st{1}{\cal T}_{-}(x,\al) r(-\al+\de,\ga) \st{1}{\cal T}_{+}(x,\al)\right]  .
\end{eqnarray*}
After taking trace on ${\cal T}$, we find
 $$  \left\{ tr\st{1}{\cal T}(x,\al),tr\st{2}{\cal T}(x,\ga) \right\}
 =tr_{1}tr_{2}  \left\{ \st{1}{\cal T}(x,\al),\st{2}{\cal T}(x,\ga) \right\}
 =0  , $$
that is
\begin{eq}
  \left\{ \st{1}{\tau}(\al),\st{2}{\tau}(\ga) \right\}=0   . \ll{0t}
\end{eq}
\indent
In other words, ${\tau(\al)}$ constructs a one-parameter involutive family. Here we remark that no symmetry conditions of 
$r$- matrix is used to obtain equation (\ref{0t}). Consequently it can be applied to general model.

\vspace{-5mm}
\subsection{Conservation condition}
\vspace{-5mm}
 If $\tau(\al)$ is a generating function for integrals of motion,
it must be independent of time. We find
\begin{eqnarray}
 {} & {} & \ptl_{t} \; tr{\cal T}(x,\al)       \nn \\
 &=& \ptl_{t} \; tr\left\{ K_{+}(t,\al)T(x_{+},x_{-},t,\al)K_{-}(t,\al)
              T^{-1}(x_{+},x_{-},t,-\al+\de)\right\} \nn  \\
 &=&tr\left\{[\ptl_{t}K_{+}(t,\al)-V(x_{+},t,-\al+\de)K_{+}(t,\al)+K_{+}(t,\al)
         V(x_{+},t,\al)] \right. \nn \\
  & {} &  \quad\quad \quad \times     T(x_{+},x_{-},t,\al)K_{-}(t,\al)T^{-1}(x_{+},x_{-},t,-\al+\de) \nn \\
 &{} &\quad +[\ptl_{t}K_{-}(t,\al)-V(x_{-},t,\al)K_{-}(t,\al)+K_{-}(t,\al)V(x_{-},t,-\al+\de)]   \nn \\
  & {} & \quad\quad \quad \left.  \times T^{-1}(x_{+},x_{-},t,-\al+\de)K_{+}(t,\al)T(x_{+},x_{-},t,\al) \right\}   . \ll{pt}
\end{eqnarray}
Taking $\ptl_{t}tr{\cal T}(x,\al)=0$, and supposing there has no connection between boundary variances on each end, we obtain 
the evolution equations of $\k$ matrices
\begin{eqnarray}
  \ptl_{t}K_{+}(t,\al)-V(x_{+},t,-\al+\de)K_{+}(t,\al)+K_{+}(t,\al)
         V(x_{+},t,\al) &= &0    , \nn \\
 \ptl_{t}K_{-}(t,\al)-V(x_{-},t,\al)K_{-}(t,\al)+K_{-}(t,\al)V(x_{-},t,-\al+\de)
      & =& 0    . \ll{kcb1}
\end{eqnarray}
\indent
For these equations, we find immediately that there have two isomorphisms between $K_{+}$ and $K_{-}$, 
which are $K_{+}(\al)\rightarrow K_{-}(-\al+\de)$ and $K_{+}(\al)\rightarrow K_{-}^{-1}(\al)$.  \\
\indent
If $\k$ are constant matrices ($\ptl_{t} K_{\pm}=0$), the equation (\ref{kcb1})  can be simplified as
\begin{eqnarray}
   V(x_{+},t,-\al+\de)K_{+}(t,\al) &= &K_{+}(t,\al)V(x_{+},t,\al)    , \nn \\
    V(x_{-},t,\al)K_{-}(t,\al)& = &K_{-}(t,\al)V(x_{-},t,-\al+\de)    . \ll{kcb2}
\end{eqnarray}
In the present case, we remark that $\k$ matrices are not singular matrices, so that determinants of $V$ satisfy
\begin{eq}
    \det V(x_{\pm},t,-\al+\de) = \det V(x_{\pm},t,\al)  , \ll{det}
\end{eq}
by which we can obtain the value of $\de$. After inserting $\de$ value into equation (\ref{kcb2}), we can find some
nontrival {\cic} when nontrivial $\k$ matrices appear. In other words, a class of $\k$ matrices is relate to a class of 
integrable boundary condition.  \\ 
\indent
We must point out that $\k$ matrices depending on field variables take its meaning in fact \cite{chen}.
On one hand, we must use such $\k$ matrices in order that quantity (\ref{tau1}) can be regarded as a generating 
function on periodic boundary condition too. On the other hand, studying such $\k$ matrices, we understand 
integrable condition more deeply.  \\    
\indent 
When Sklyanin's function (\ref{sk-tau}) is regarded as a generating function in {\af}, boundary $\k$ matrices will have no
constant solution in equation (\ref{kcb1}) (except for Sine-Gordon theory). So we have to solve equation (\ref{kcb1}) as 
differential equations. Besides these difficulty, even when one had found a
nontrivial solution, he should have to prove the
involution condition again because condition (\ref{ck}) may be broken.
In our method, $\de$ added on spectral parameter guarantees the existence of constant $\k$ matrices,
and $\de$ can be solved by means of equation (\ref{det}), so $\k$ matrices solving procedure is simplified effectively.  

\indent
{\bf Proposition 3.} If $\k$ matrices satisfy not only algebra equation (\ref{ckp1}) and (\ref{ckm1}), but also evolution 
equations of (\ref{kcb1}). $ tr {\cal T}(x,\al)$ is a {\gf} for integrals of motion.  

\indent
 There has a difficult step in proving this proposition. Are solutions of $\k$ matrices in equation (\ref{kcb1}) compatible 
with its algebra equation (\ref{ckp1}) and (\ref{ckm1}) ? Although one found it is true in Sine-Gordon theory \cite{M} and 
it has been proved in {\af} \cite{BCD} with the form of (\ref{tau2}), it still keeps an open problem in general theory.
If this compatibility is satisfied, the proposition is proved naturally.

\vspace{-5mm}
\subsection{Other generating functions}
\vspace{-5mm}
As exhibited in above subsections, we also obtain other $\k$ matrices' algebra and evolution equations when quantity (\ref{tau2}) 
or (\ref{tau3}) is regarded as generating function for integrals of motion. In the form of (\ref{tau2}), it read as
\begin{eqnarray}
   0  &=&  r^{t_{1}t_{2}}(-\al+\de,-\ga+\de) {\st{1}{K}}_{+}(t,\al) {\st{2}{K}}_{+}(t,\ga)
    +{\st{1}{K}}_{+}(t,\al) r^{t_{2}}(\al,-\ga+\de)  {\st{2}{K}}_{+}(t,\ga)     \nn \\
 {}&{} & + {\st{2}{K}}_{+}(t,\ga) r^{t_{1}}(-\al+\de,\ga) {\st{1}{K}}_{+}(t,\al)
    + {\st{1}{K}}_{+}(t,\al) {\st{2}{K}}_{+}(t,\ga) r(\al,\ga)      , \ll{classical} \\
  0  &=&      r(\al,\ga)    \st{1}{K}_{-}(t,\al) \st{2}{K}_{-}(t,\ga)
     +\st{1}{K}_{-}(t,\al) r^{t_{1}}(-\al+\de,\ga)    \st{2}{K}_{-}(t,\ga)  \nn \\
 {}&{} & +  \st{2}{K}_{-}(t,\ga) r^{t_{2}}(\al,-\ga+\de)     \st{1}{K}_{-}(t,\al)
   +  \st{1}{K}_{-}(t,\al) \st{2}{K}_{-}(t,\ga) r^{t_{1}t_{2}}(-\al+\de,-\ga+\de)  . \nn
\end{eqnarray}
where the upper index "$t_{i}, i=1,2$" denote transposition on the "$ i $" space.
its evolution equations read as
\begin{eqnarray}
   0 &=& \ptl_{t}K_{+}(t,\al)+V^{t}(x_{+},t,-\al+\de)K_{+}(t,\al)+K_{+}(t,\al)
         V(x_{+},t,\al)    , \nn \\
  0 &=& \ptl_{t}K_{-}(t,\al)-V(x_{-},t,\al)K_{-}(t,\al)-K_{-}(t,\al)V^{t}(x_{-},t,-\al+\de)   .  \ll{delta2}
\end{eqnarray}
\indent
Let $\ptl_{t} \k =0$, we can also obtain $\de$ value in (\ref{delta2})
by taking determinants. With constraint conditions of (\ref{classical}) and (\ref{delta2}),
quantity (\ref{tau2}) is a generating function for integrals of motion. \\
\indent
In the form of (\ref{tau3}), the similar equations are still balance except
that we must use "$\dag$" (conjugation) instead of "$t$"
(transposition). If those modified equations are satisfied, quantity
(\ref{tau3}) can be regarded as a {\gf } for integrals of motion too. \\
\indent
Now, we have obtained three forms of {\gf} as well as their constraint conditions.
If all of them are regarded as {\gf s}, we believe that they are the same one in fact.
In the next section, we will prove it explicitly in {\af}.

\section{Classically Integrable boundary condition in {\af}}
\vspace{-5mm}
\subsection{Links among {\gf s}}
\vspace{-5mm}
Lagrangian in {\af} on independent boundary condition is \cite{{BCD},{MOP}}:
\begin{eqnarray}
   {\cal L}=\ii dx \ii & dt & \left\{ \th (x -x_{-})\th (x_{+}-x)[\frac{1}{2}\ptl_{\mu}\phi_{a} \ptl^{\mu}\phi_{a}
         -\frac{m^{2}}{\be^{2}}\sum\limits_{0}^{r} n_{i} (e^{\be\al_{i}\cdot \phi}-1)]  \right.   \nn \\
 {} &{} & \left. -\de (x-x_{-}) V_{-}(\phi(x_{-}),\ptl_{\mu}\phi(x_{-}))
                 -\de (x_{+}-x) V_{+}(\phi(x_{+}),\ptl_{\mu}\phi(x_{+})) \right  \}   , \nn \\     
 && \ll{afl}
\end{eqnarray}
where $m$ is mass scale and $\be$ is the coupling constant in real domain; $\al_{i}$ are simple roots of a simple Lie algebra 
of rank $r$ (included the affine root $\al_{0}$). There are $\sum\limits_{0}^{r} n_{i}\al_{i}=0$ and $n_{0}=1$.
It is a theory of $r$ scalar fields ($\al_{i}\cdot \phi=\sum\limits_{a=0}^{r-1} \al_{i}^{a}\phi_{a}$).
The potentials $V_{+}$ and $V_{-}$ are additions on the ends $x_{+}$ and
$x_{-}$, respectively. They denote independent boundary conditions.
When $V_{\pm}$ depend on $\phi(x_{\pm})$ only (and independent of its derivatives), we obtain:
\begin{eqnarray}
   (\ptl^{2}_{t}-\ptl^{2}_{x})\phi &=& -\frac{m^{2}}{\be}\sum\limits_{0}^{r} n_{i}\al_{i} e^{\be\al_{i}\cdot \phi}  ,
                                                                                            \quad x_{-}< x <x_{+}   , \nn \\
   \ptl_{x} \phi_{a} &=& \mp\frac{\ptl V_{\pm}}{\ptl \phi_{a}}  , \kog \quad\quad x=x_{\pm}    .     \ll{bound-eq}
\end{eqnarray}
\indent 
Lax pair in {\af} read as ($\la=e^{\al}$) 
\begin{eqnarray}
  U(x,t,\la)&=& -\left\{ \frac{1}{2}\be H\cdot \ptl_{t}\phi +m \sum\limits_{0}^{r}
      \sqrt{m_{i}}(\la E_{\al_{i}}+\la^{-1} E_{-\al_{i}}) e^{\be\al_{i}\cdot \phi/2} \right\}   , \nn \\
  V(x,t,\la)&=& -\left\{ \frac{1}{2}\be H\cdot \ptl_{x}\phi +m \sum\limits_{0}^{r}
      \sqrt{m_{i}}(\la E_{\al_{i}}-\la^{-1} E_{-\al_{i}}) e^{\be\al_{i}\cdot \phi/2} \right\}   . \ll{auv}
\end{eqnarray}
in which $H$ and $E_{\pm \al_{i}}$ are the Cartan subalgebra and the generators responding to the simple roots, respectively, 
of the simple Lie algebra of rank $r$. The coefficients $m_{i}$ are equal to $n_{i} \al_{i}^{2}/8$. There are the Lie algebra 
relation:
 $$ [H_{i},H_{j}] = 0  \quad ,\;[H,E_{\pm\al_{i}}]=\pm\al_{i} E_{\pm\al_{i}}   , $$
\begin{eq}
 \left[E_{\al_{i}},E_{-\al_{i}}\right] = 2 \al_{i}\cdot H/(\al_{i}^{2})  . \ll{lie}
\end{eq}
\indent
It is pointed out by Hollowood \cite{H} that the complex affine Toda theories have soliton solution (in which coupling constant 
$\be$ is purely imaginary), in contrast with the real coupling constant ones. And its Lagrangian, motion equation and Lax pair 
can be expressed similarly to equation (\ref{afl})-(\ref{auv}) except for taking $\be\rightarrow i\wt{\be}$ ($\wt{\be}\in Re$). 
In our paper, we use (\ref{afl})-(\ref{auv}) equations in general, and distinguish them only when the real and imaginary cases
can't be treated in the same way. \\
\indent
 For those generators in Lax pair (\ref{auv}), we can find a representation in which they satisfy:
 $$ H_{i}^{t}=H_{i}^{\dag}=H_{i} \;\;
        ,E_{\pm\al_{i}}^{t}=E_{\pm\al_{i}}^{\dag}=E_{\mp\al_{i}}  . $$
So there has an automorphism map :
\begin{eqnarray*}
  H_{i}\rightarrow {H^{\prime}}_{i} &=& \Om^{-1} H_{i} \Om =  -H_{i} \; , \\
  E_{\al_{i}}\rightarrow {E^{\prime}}_{\al_{i}} &=& \Om^{-1} E_{\al_{i}} \Om = E_{-\al_{i}}  \; , \\
  E_{-\al_{i}}\rightarrow {E^{\prime}}_{-\al_{i}} &=& \Om^{-1} E_{-\al_{i}} \Om = E_{\al_{i}} \; .
\end{eqnarray*}
The new generators satisfy the same Lie algebra relation (\ref{lie}). In other words, there are
\begin{ea}
 & U^{t}(x,\la) =U(x,\la^{-1}) = -\Om^{-1}U(x,-\la)\Om  \; , & \nn \\
 & V^{t}(x,\la) =V(x,-\la^{-1})= -\Om^{-1}V(x,-\la)\Om   \; . &\ll{uvt}
\end{ea}
We remark that equation (\ref{uvt}) can be applied to both real and imaginary
coupling constant cases. But if one uses "$\dag$" instead of "$t$", equation
(\ref{uvt}) must be modified because of its complex fields. \\
\indent
 From definition of $T(x,y,t,\la)$, there has
\begin{eqnarray*}
      \ptl_{x}T^{t}(x,y,\la) &=& T^{t}(x,y,\la)U^{t}(x,\la)   \\
                          &=& T^{t}(x,y,\la)\left[ -\Om^{-1}U(x,-\la)\Om\right]   ,
\end{eqnarray*}
or
$$ \ptl_{x}\left[\Om T^{t}(x,y,\la) \Om^{-1}\right]
            = -\left[\Om T^{t}(x,y,\la) \Om^{-1}\right]U(x,-\la)  . $$  
Comparing it with $\ptl_{x}T^{-1}(x,y,\la) = -T^{-1}(x,y,\la)U(x,\la)$ and the initial condition in (\ref{tm}), we obtain
  $$ \Om T^{t}(x,y,\la) \Om^{-1}=T^{-1}(x,y,-\la)  ,$$
or
\begin{eq}
   \Om T^{t}(x,y,\al) \Om^{-1}=T^{-1}(x,y,\al+i\pi)  .
\end{eq}
It means
\begin{eqnarray}
 & {} &tr\left\{K_{-}(\al)T^{-1}(-\al+\de)K_{+}(\al)T(\al)\right\}        \nn \\
  &=& tr\left\{K_{-}(\al)\Om T^{t}(-\al+\de+i\pi)\Om^{-1} K_{+}(\al)T(\al)\right\}    \nn \\
   &  =& tr\left\{\wt{K}_{-}(\al)T^{t}(-\al+\de^{\prime})\wt{K}_{+}(\al)T(\al)\right\}  .\ll{trt}
\end{eqnarray}
in which $\k$ matrices in (\ref{tau1}) and (\ref{tau2}) are distinguished by $\k$ and $\wt{K}_{\pm}$ now, 
and the quantities added on spectral parameter become $\de$ and $\de^{\prime}$ respectively.
In other words, quantity (\ref{tau1}) is equal to (\ref{tau2}), if
$\de^{\prime}$ is equal to $\de+i\pi$ and reflection matrices satisfy
\begin{eq}
   \wt{K}_{-}(\al)=K_{-}(\al)\Om  \; , \;  \wt{K}_{+}(\al)=\Om^{-1} K_{+}(\al)  . \ll{wtk1}
\end{eq}
\indent
Using the second equation of (\ref{uvt}) and comparing equation (\ref{kcb1})
with (\ref{delta2}), we find these relations appear again. So quantities
(\ref{tau1}) and (\ref{tau2}) are the same one in fact, when both of them
are regarded as {\gf s} for integrals of motion. \\
\indent
 In real coupling constant and real fields case, if we use "$\dag$" instead of "$t$", equation (\ref{uvt}) is still balance 
when spectral parameter is real. So we obtain a relation similar to (\ref{trt}) again. 
In this case, when we rewrite (\ref{tau3}) as 
$\tau(\al)=tr \bar{K}_{-}(\al)T^{\dag}(-\al+\de^{\prime\prime}){\bar{k}_{+}(\al)T(\al)}$, then
\begin{eq}
 \wt{K}_{\pm}(\al)=\bar{K}_{\pm}(\al) \quad ,\; \de^{\prime}=\de^{\prime\prime}=\de+i\pi  . \ll{wtk2}
\end{eq}
\indent
 Now, we have proved quantity (\ref{tau1}) is equal to (\ref{tau2}) when both of them are regarded as generating
functions in {\af}. When coupling constant is real, they are equal to {\gf} (\ref{tau3}) too. But when coupling
constant is purely imaginary, equation (\ref{uvt}) may be not satisfied, so $\k$ matrices in (\ref{tau3})
may have no constant solution.
 
\vspace{-5mm}
\subsection{Classically Integrable Boundary Condition}
\vspace{-5mm}
 In real coupling constant {\af}, if we regard $\tau(\al)=tr\bar{K}_{-}(\al)T^{\dag}(-\al+\de){\bar{K}_{+}(\al)T(\al)}$
as a {\gf}, we will obtain the evolution equation of $\bar{K}_{+}$ in $x_{+}$ boundary:
\begin{eq}
  V^{\dag}(x_{+},-\al+\de)\bar{K}_{+}(\al)+\bar{K}_{+}(\al)V(x_{+},\al)=0  .
\end{eq}
After taking $\de=0$, we obtain equation of $\bar{K}^{-1}_{+}(\la)$(in which $\la=e^{\al}$):
\begin{eq}
 \frac{1}{2}\left[ \bar{K}^{-1}_{+}(\la),\frac{\be}{m}\ptl_{x}\phi \cdot H\right]_{+}=
    \left[\bar{K}^{-1}_{+}(\la),\sum\limits_{0}^{r}\sqrt{m_{i}}(\la E_{\al_{i}}-\la^{-1}E_{-\al_{i}})e^{\be\al_{i}\cdot \phi/2}
    \right]_{-} \; \; . \ll{pbc}
\end{eq}
\indent By boundary equation (\ref{bound-eq}), it is just the reflection equation appears in paper \cite{BCD}. 
We find $K_{+}$ and $T$ matrices defined in that paper are just the quantities of $\bar{K}^{-1}_{+}$ and ${T}^{-1}$ 
in our paper, according to different definition of Lax pair. Analogy with method in paper \cite{BCD},
we solve equation (\ref{pbc}) and obtain {\cic} in {\af}. In simple-laced case, it is the same as paper \cite{BCD}
\begin{eqnarray}
 \frac{\be}{m}\ptl_{x}\phi &=& -\sum\limits_{0}^{r} B_{i}\sqrt{\frac{n_{i}}{2|\al_{i}|^{2}}}
          \al_{i} e^{\be\al_{i}\cdot\phi/2} ,      \nn \\
 \hbox{in which}\quad\; |B_{i}| &=&2 \;\;\; ,i=0,1,\cdots,r ,  \nn \\
 \hbox{or} \;\quad ,B_{i}&=&0\quad, i=0,1,\cdots,r   . 
\end{eqnarray}
\indent
 In imaginary coupling constant one, we regard $\tau(\al)=tr\wt{K}_{-}(\al)T^{t}(-\al+\de){\wt{K}_{+}(\al)T(\al)}$ 
as a {\gf}. The results in real coupling constant can be used, thanks to section 4.1. In other words, the new {\cic} can 
be obtained by analytic continuation by $\be \rightarrow i\wt{\be}$  ($\wt{\be}\in Re$). It reads:
\begin{eqnarray}
 \frac{i\wt{\be}}{m}\ptl_{x}\phi &=& -\sum\limits_{0}^{r} B_{i}\sqrt{\frac{n_{i}}{2|\al_{i}|^{2}}}
          \al_{i} e^{i\wt{\be}\al_{i}\cdot\phi/2}  ,         \nn \\
  \hbox{in which} \;\quad |B_{i}| &=&2 \;\;\; ,i=0,1,\cdots,r ,  \nn \\
 \hbox{or} \quad\; B_{i}&=&0\quad, i=0,1,\cdots,r   . 
\end{eqnarray}
\indent
 We remark that Sklyanin's method \cite{S1} can't be used in {\af} except for Sine-Gorden theory, this conclusion comes 
from the fact that $\de\not=0$ on independent boundary condition if we regard (\ref{tau1}) as a {\gf}. Now, we must take 
$\de=-i\pi$ according to equation (\ref{wtk1}) and (\ref{wtk2}). Sine-Gordon theory is an exception in which it is 
satisfied both $\de=0$ and $\de=-i\pi$. \\
\indent 
As we discussed in section 3, in the classical case, no symmetry conditions
of $r$- matrix is necessary in constructing generating function for integrals
of motion. So it is interesting to study whether commutative quantities can be constructed with less symmetry of
$R$- matrix in the quantum case.

\section{Quantum integrable systems with boundary}
There are many papers (for example, \cite{S1} and \cite{MN} - \cite{Kulish}) in which
the authors deal with quantum integrable boundary condition
in two-dimensional lattice models. As far as we know, both unitarity and crossing unitarity conditions 
(or the weaker property \cite{Re}) are used in constructing commutative quantities. Since finding crossing unitarity condition 
of a given $R$- matrix is a difficult problem, it is useful to construct commutative quantities without this symmetry.\\
\indent
 In this section, we explore how to obtain commutative quantities by means of unitarity condition only.
Unitarity condition read as:
\begin{eq}
  R_{12}(u)R_{21}(-u)=\xi(u)  ,    \ll{uc}
\end{eq}  
where $\xi(u)$ is some even scalar function and $R$- matrix is a solution of quantum Yang-Baxter equation (YBE):
  $$ R_{12}(u-v)R_{13}(u)R_{23}(v)=R_{23}(v)R_{13}(u)R_{12}(u-v)   .$$
\indent 
As usual, the transfer matrix $t(u)$ is defined as
\begin{eq}
 t(u)= tr \; {\cal T}_{+}(u) {\cal T}_{-}(u)  , \ll{qt}
\end{eq}  
and each entry of ${\cal T}_{+}(u)$ commutes with ${\cal T}_{-}(u)$.

\vspace{0.4mm}
\indent
{\bf Proposition 4.} If ${\cal T}_{\pm}$ satisfy such equations
\begin{eqnarray}
      R_{21}^{t_{1}t_{2}}(-u_{-}) \st{1}{\cal T}_{+}^{t_{1}}(u) R_{21}^{t_{2}}(u_{+}-\de) \st{2}{\cal T}_{+}^{t_{2}}(v) &=&
     \st{2}{\cal T}_{+}^{t_{2}}(v) R_{12}^{t_{1}}(u_{+}-\de) \st{1}{\cal T}_{+}^{t_{1}}(u) R_{12}(-u_{-})           ,  \nn  \\
      R_{12}(u_{-}) \st{1}{\cal T}_{-}(u) R_{12}^{t_{1}}(-u_{+}+\de)  \st{2}{\cal T}_{-}(v)  &=&
     \st{2}{\cal T}_{-}(v) R_{21}^{t_{2}}(-u_{+}+\de) \st{1}{\cal T}_{-}(u)  R_{21}^{t_{1}t_{2}}(u_{-})            ,   \ll{te}
\end{eqnarray}
and the quantum $R$- matrix obeys the unitarity condition, then, transfer matrix $t(u)$ defines a one-parameter commutative family.

\vspace{0.4mm}
\indent
For explicity, we use $\xi_{1}^{-1}$ and $\xi_{2}^{-1}$ replace
$\xi^{-1}(u_{+}-\de)$ and $\xi^{-1}(-u_{-})$ respectively,
as well as $u_{\pm}=u\pm v$. The proof is directly: 
\begin{eqnarray*}
  t(u)t(v) &=& tr_{1} \st{1}{\cal T}_{+}(u) \st{1}{\cal T}_{-}(u)  \; tr_{2} \st{2}{\cal T}_{+}(v) \st{2}{\cal T}_{-}(v)  \\   
             &=& tr_{12} \st{1}{\cal T}_{+}^{t_{1}}(u) \st{2}{\cal T}_{+}(v) \st{1}{\cal T}_{-}^{t_{1}}(u) \st{2}{\cal T}_{-}(v) \\
             &=& \xi_{1}^{-1} \; tr_{12} \st{1}{\cal T}_{+}^{t_{1}} \st{2}{\cal T}_{+} 
                 R_{21}(u_{+}-\de)R_{12}(-u_{+}+\de) \st{1}{\cal T}_{-}^{t_{1}} \st{2}{\cal T}_{-} \\
             &=& \xi_{1}^{-1} \; tr_{12} \{\st{1}{\cal T}_{+}^{t_{1}} R_{21}^{t_{2}}(u_{+}-\de) \st{2}{\cal T}_{+}^{t_{2}} \}^{t_{2}}
                \{ \st{1}{\cal T}_{-} R_{12}^{t_{1}}(-u_{+}+\de)  \st{2}{\cal T}_{-} \}^{t_{1}} \\
             &=& \xi_{1}^{-1}\xi_{2}^{-1} \;
                 tr_{12} \{\st{1}{\cal T}_{+}^{t_{1}} R_{21}^{t_{2}}(u_{+}-\de) \st{2}{\cal T}_{+}^{t_{2}} \}^{t_{1}t_{2}}
                R_{21}(-u_{-})R_{12}(u_{-})  \{ \st{1}{\cal T}_{-} R_{12}^{t_{1}}(-u_{+}+\de)  \st{2}{\cal T}_{-} \} \\
             &=& \xi_{1}^{-1}\xi_{2}^{-1} \;
                 tr_{12} \{R_{21}^{t_{1}t_{2}}(-u_{-}) \st{1}{\cal T}_{+}^{t_{1}} R_{21}^{t_{2}}(u_{+}-\de) \st{2}{\cal T}_{+}^{t_{2}} \}^{t_{1}t_{2}}
                 \{ R_{12}(u_{-}) \st{1}{\cal T}_{-} R_{12}^{t_{1}}(-u_{+}+\de)  \st{2}{\cal T}_{-} \}  ,
\end{eqnarray*}
using equations (\ref{te}), we find
\begin{eqnarray*}
   t(u)t(v)&=& \xi_{1}^{-1}\xi_{2}^{-1} \;
                 tr_{12} \{ \st{2}{\cal T}_{+}^{t_{2}} R_{12}^{t_{1}}(u_{+}-\de) \st{1}{\cal T}_{+}^{t_{1}} R_{12}(-u_{-}) \}^{t_{1}t_{2}}
                 \{  \st{2}{\cal T}_{-} R_{21}^{t_{2}}(-u_{+}+\de) \st{1}{\cal T}_{-}  R_{21}^{t_{1}t_{2}}(u_{-}) \}    \\
             &=& \xi_{1}^{-1} \;
                 tr_{12} \{ \st{2}{\cal T}_{+}^{t_{2}} R_{12}^{t_{1}}(u_{+}-\de) \st{1}{\cal T}_{+}^{t_{1}}  \}^{t_{1}t_{2}}
                 \{  \st{2}{\cal T}_{-} R_{21}^{t_{2}}(-u_{+}+\de) \st{1}{\cal T}_{-} \}    \\
             &=& \xi_{1}^{-1} \;
                 tr_{12} \{ \st{2}{\cal T}_{+}^{t_{2}} R_{12}^{t_{1}}(u_{+}-\de) \st{1}{\cal T}_{+}^{t_{1}}  \}^{t_{1}}
                 \{  \st{2}{\cal T}_{-} R_{21}^{t_{2}}(-u_{+}+\de) \st{1}{\cal T}_{-} \}^{t_{2}}    \\   
             &=& tr_{12}  \st{2}{\cal T}_{+}^{t_{2}}  \st{1}{\cal T}_{+} \st{2}{\cal T}_{-}^{t_{2}} \st{1}{\cal T}_{-}  \\
             &=& t(v)t(u) .
\end{eqnarray*}
\indent
In quantum spin chain model, it is convenient that ${\cal T}_{\pm}$ take such representations
\begin{eqnarray}
  {\cal T}_{+}(u) &=& K_{+}(u)   ,          \nn   \\
  {\cal T}_{-}(u) &=& T(u) K_{-}(u) T^{t}(-u+\de)  \ll{rep} \\
                  &=& L_{N}(u)\cdots L_{2}(u)L_{1}(u) K_{-}(u) L_{1}^{t}(-u+\de)L_{2}^{t}(-u+\de)\cdots L_{N}^{t}(-u+\de)  ,\nn  
\end{eqnarray}  
in which transposition "$t$" acts on the auxiliary space and $n=1,2,...,N$ denote quantum space. There has a relation between 
$R$ and $L$ operators
\begin{eq}
 R_{ab}(u-v)L_{a}(u)L_{b}(v)=L_{b}(v)L_{a}(u)R_{ab}(u-v)  .
\end{eq}
\indent
Let ${\cal T}_{-}(u)=L_{N}(u) {\cal T}_{-}^{\prime}(u)L_{N}^{t}(-u+\de)$ and
insert equation (\ref{rep}) into equation (\ref{te}), we find the second equation of (\ref{te}) becomes
\begin{eqnarray*}
  l.h.s. &=& R_{ab}(u_{-}) L_{aN}(u) \st{a}{{\cal T}_{-}^{\prime}} L_{aN}^{t_{a}}(-u+\de) 
             R_{ab}^{t_{a}}(-u_{+}+\de) L_{bN}(v) \st{b}{{\cal T}_{-}^{\prime}} L_{bN}^{t_{b}}(-u+\de) \\
         &=& R_{ab}(u_{-}) L_{aN}(u) \st{a}{{\cal T}_{-}^{\prime}} L_{bN}(v)  
             R_{ab}^{t_{a}}(-u_{+}+\de) L_{aN}^{t_{a}}(-u+\de) \st{b}{{\cal T}_{-}^{\prime}} L_{bN}^{t_{b}}(-u+\de) \\
         &=& L_{bN}(v) L_{aN}(u) R_{ab}(u_{-}) \st{a}{{\cal T}_{-}^{\prime}} 
             R_{ab}^{t_{a}}(-u_{+}+\de)  \st{b}{{\cal T}_{-}^{\prime}} L_{aN}^{t_{a}}(-u+\de) L_{bN}^{t_{b}}(-u+\de)  ,
\end{eqnarray*}
\begin{eqnarray*}
  r.h.s. &=& L_{bN}(v) \st{b}{{\cal T}_{-}^{\prime}} L_{bN}^{t_{b}}(-u+\de) R_{ba}^{t_{b}}(-u_{+}+\de)
             L_{aN}(u) \st{a}{{\cal T}_{-}^{\prime}} L_{aN}^{t_{a}}(-u+\de) R_{ba}^{t_{a}t_{b}}(u_{-})  \\
         &=& L_{bN}(v) \st{b}{{\cal T}_{-}^{\prime}} L_{aN}(u) R_{ba}^{t_{b}}(-u_{+}+\de)
             L_{bN}^{t_{b}}(-u+\de) \st{a}{{\cal T}_{-}^{\prime}} L_{aN}^{t_{a}}(-u+\de) R_{ba}^{t_{a}t_{b}}(u_{-})  \\
         &=& L_{bN}(v) L_{aN}(u) \st{b}{{\cal T}_{-}^{\prime}} R_{ba}^{t_{b}}(-u_{+}+\de)
             \st{a}{{\cal T}_{-}^{\prime}} R_{ba}^{t_{a}t_{b}}(u_{-}) L_{aN}^{t_{a}}(-u+\de)  L_{bN}^{t_{b}}(-u+\de)  . \\
\end{eqnarray*}
In other words, this equation is reduced to
$$  R_{ab}(u_{-}) \st{a}{{\cal T}_{-}^{\prime}} R_{ab}^{t_{a}}(-u_{+}+\de)  \st{b}{{\cal T}_{-}^{\prime}} = 
    \st{b}{{\cal T}_{-}^{\prime}} R_{ba}^{t_{b}}(-u_{+}+\de) \st{a}{{\cal T}_{-}^{\prime}} R_{ba}^{t_{a}t_{b}}(u_{-})    . $$    
\indent
We proceed to do the above reduction repeatedly until all of $L$ operators beside $K_{-}$ matrix disappear. At last, we obtain the
reflection equation about $K_{-}$ only. Now, reflection equations of $\k$ are
\begin{eqnarray}
      R_{21}^{t_{1}t_{2}}(-u_{-}) \st{1}{K}_{+}^{t_{1}}(u) R_{21}^{t_{2}}(u_{+}-\de) \st{2}{K}_{+}^{t_{2}}(v) &=&
     \st{2}{K}_{+}^{t_{2}}(v) R_{12}^{t_{1}}(u_{+}-\de) \st{1}{K}_{+}^{t_{1}}(u) R_{12}(-u_{-})                   ,  \nn  \\
      R_{12}(u_{-}) \st{1}{K}_{-}(u) R_{12}^{t_{1}}(-u_{+}+\de)  \st{2}{K}_{-}(v)  &=&
     \st{2}{K}_{-}(v) R_{21}^{t_{2}}(-u_{+}+\de) \st{1}{K}_{-}(u)  R_{21}^{t_{1}t_{2}}(u_{-})           ,      \ll{k-re}
\end{eqnarray}
and transfer matrix $t(u)$ becomes
\begin{eq}
 t(u)= tr \; K_{+}(u) T(u) K_{-}(u) T^{t}(-u+\de) \; . \ll{qt1}
\end{eq} 
\indent
We remark that there has no obvious relation between $\k$ matrices.
If some symmetry conditions are used, relation between $K_{+}$ and $K_{-}$
matrices may be found. \\
\indent
For example, $R$- matrix in paper \cite{MN} has $PT$ symmetry and crossing unitarity
\begin{eqnarray}
   R_{12}(u) &=& R_{21}^{t_{1}t_{2}}(u)      , \nn   \\
   R_{12}(u) &=& \st{1}{V} R_{12}^{t_{2}}(-u-\rho) \st{1}{V}^{-1}   . \ll{cuc}
\end{eqnarray}
\indent
By $PT$ symmetry, we find there has an isomorphism between boundary matrices:
\begin{equation}
   K_{-}(u)=K_{+}^{t}(-u+\de).  \ll{rer1}
\end{equation}
If both $PT$ symmetry and crossing unitarity are considered, there is another relation as
\begin{equation}
  K_{-}(u)=K_{+}^{-1}(u+\rho) M^{-1} , \quad \quad M=V^{t} V. \ll{rer2}
\end{equation}
From (\ref{rer1}) and (\ref{rer2}), it means
\begin{equation}
  K_{+}^{t}(-u+\de) M K_{+}(u+\rho)= Id , \quad
  K_{-}^{t}(-u+\de) K_{-}(u-\rho) M =Id .
\end{equation}
These equations may be regarded as constraint conditions on $\de$. \\
\indent
Now, it is interesting to compare our commutative quantities with those
of Mezincescu and Nepomechie \cite{MN}. Using conditions of (\ref{cuc})
and unitarity condition $R_{12}(u)R_{21}(-u)=\xi(u)$, we obtain
\begin{eqnarray*}
  R_{12}^{t_{1}}(-u-\rho) &=& (\st{1}{V})^{t_{1}} \; R_{12}^{t_{1}t_{2}}(u)  \; (\st{1}{V}^{-1})^{t_{1}}                  \\
                    &=& \xi(-u) \; (\st{1}{V})^{t_{1}} \;   R_{12}^{-1}(-u) \; (\st{1}{V}^{-1})^{t_{1}}   ,
\end{eqnarray*}
or
$$
  R_{12}^{t_{1}}(-u+\de)
   = \xi(-u+\rho+\de) \; (\st{1}{V})^{t_{1}} \;
    R_{12}^{-1}(-u+\rho+\de) \; (\st{1}{V}^{-1})^{t_{1}}   .
$$
\indent
If $L_{n}(u)$ is defined as $L_{n}(u)\equiv L_{an}(u)=R_{an}(u)$, we obtain
\begin{eqnarray*}
 L_{n}^{t}(-u+\de) &=& \xi(-u+\rho+\de) \; {V}^{t}  \; L_{n}^{-1}(-u+\rho+\de)  \;  ({V}^{-1})^{t}  , \\
 T^{t}(-u+\de)  &=& L_{1}^{t}(-u+\de) \;  L_{2}^{t}(-u+\de) \cdots L_{N}^{t}(-u+\de)  \\
           &=&\xi^{N}(-u+\rho+\de) \; {V}^{t} \;  T^{-1}(-u+\rho+\de)  \;  ({V}^{-1})^{t}    .
\end{eqnarray*} 
In other words, transfer matrix (\ref{qt1}) becomes
\begin{eqnarray} 
  t(u)&=& tr\;  K_{+}(u) T(u) K_{-}(u) \;  (\xi^{N}(-u+\rho+\de) \; {V}^{t} T^{-1}(-u+\rho+\de) ({V}^{-1})^{t})    \nn \\
      &=& \xi^{N}(-u+\rho+\de) \;tr\;  (({V}^{-1})^{t} K_{+}(u)) \; T(u) \; (K_{-}(u) {V}^{t}) \; T^{-1}(-u+\rho+\de)   .  \ll{comp}
\end{eqnarray}  
\indent
It is convenient to multiply (\ref{comp}) by $\xi^{-N}(-u+\rho+\de)$ before
we regard it as the commutative quantities, {\it i. e.},
 $$ t(u)=  tr\; K_{+}^{\prime}(u) T(u) K_{-}^{\prime}(u) T^{-1}(-u+\rho+\de). $$
If $\de$ is equal to $-\rho$, it is just the one in paper \cite{MN}. \\
\indent
As one of the main results in our paper, we have constructed the commutative
quantities with unitarity condition of quantum $R$- matrix only. As discussed
in the classical case, with symmetry conditions of (\ref{cuc}), we can find
another form of commutative quantities, and these two forms are the same one
in fact when both of them are regarded as commutative quantities. \\
\indent
Finally, we study the classical counterparts of reflection equations
(\ref{k-re}) by modifying unitarity condition to $R_{12}(u)R_{21}(-u)=Id$.
In the classical limit, as $\hbar\rightarrow 0$, one
has \cite{{S1},{Faddeev}}:
$$ [ \; , \; ]= -i \hbar \{\; , \; \} \; ;  \quad R(u)=Id+i \hbar \; r(u) + o({\hbar}^{2})  . $$
So unitarity condition means $r_{12}(u)=-r_{21}(-u)$ and quantum YBE goes over into the classical YBE.
We find reflection equations (\ref{k-re}) just turn into equation (\ref{classical}) in which
$r(\al,\be)$ is equal to $r(\al-\be)$ now. In contrast with
the quantum case, there has an isomorphism between $K_{+}$ and $K_{-}$, 
which is $K_{+}(\al)\rightarrow K_{-}^{-1}(\al)$.
 
\section{Conclusion and discussion}
 In this paper, we obtain three possible generating functions for integrals
of motion in classically integrable field theory on a finite interval with independent
boundary condition on each end. As constraint conditions, we find $\k$ matrices'
algebra and evolution equations. In contrast with other's methods, a new parameter
is added on spectral parameter, and we expect it shall simplify the procedure
of solving $\k$ matrices effectively.  In {\af}, we prove these generating functions
are equivalent to each other and its links are discussed too. Our results show
that two of these generating functions are always valid in both real and imaginary
coupling constant cases.  \\
\indent
It is remarkable that no symmetry condition of $r$- matrix is used when
we regard quantities (\ref{tau1}), (\ref{tau2}) and (\ref{tau3}) as generating
functions for integrals of motion, so we expect it shall be applied to more integrable models than
\cite{{S1},{BCD}}. As demonstrated in section 4, the added parameter $\de$
improves this possibility. \\
\indent
We also extend our results to quantum spin chain, we have proved that unitarity
condition of quantum $R$- matrix is sufficient to construct commutative quantities
with boundary. Reflect equations of $\k$ are obtained. Relation between
boundary $\k$ matrices is found when $PT$ symmetry and crossing
unitarity condition of $R$- matrix are considered.
With these symmetry conditions, we also find another form of commutative quantities
as the one defined in \cite{MN}. Finally, we find that
classical counterparts of quantum reflection equations is just the one
which we obtain by classical quantity (\ref{tau2}).

\section*{Acknowledgement}
This work was supported by Climbing Up Project, NSCC, Natural 
Scientific Foundation of Chinese Academy of Sciences and 
Foundation of NSF.

\end{document}